\documentclass[prl,aps,twocolumn,superscriptaddress]{revtex4-1}

\usepackage{color}
\usepackage{tabularx}
\usepackage{lineno}
\usepackage{hyperref}
\hypersetup {
pdfpagemode = {UseNone},
pdftitle = {Research towards high-repetition rate laser-driven X-ray sources for imaging applications},
pdfauthor = {Johannes Götzfried}
} 
\usepackage{xcolor}
\usepackage{siunitx}
\usepackage{upgreek}
\usepackage[figuresright]{rotating}

\begin{document}

\title{Research towards high-repetition rate laser-driven X-ray sources for imaging applications}

\author{J. G\"otzfried}
\affiliation{Ludwig-Maximilians-Universit\"at M\"unchen, Am Coulombwall 1, 85748 Garching, Germany}

\author{A. D\"opp}
\affiliation{Ludwig-Maximilians-Universit\"at M\"unchen, Am Coulombwall 1, 85748 Garching, Germany}

\author{M. Gilljohann}
\affiliation{Ludwig-Maximilians-Universit\"at M\"unchen, Am Coulombwall 1, 85748 Garching, Germany}

\author{H. Ding}
\affiliation{Ludwig-Maximilians-Universit\"at M\"unchen, Am Coulombwall 1, 85748 Garching, Germany}

\author{S. Schindler}
\affiliation{Ludwig-Maximilians-Universit\"at M\"unchen, Am Coulombwall 1, 85748 Garching, Germany}

\author{J. Wenz}
\affiliation{Ludwig-Maximilians-Universit\"at M\"unchen, Am Coulombwall 1, 85748 Garching, Germany}

\author{L. Hehn}
\affiliation{Lehrstuhl f\"ur Biomedizinische Physik, Physik-Department \& Munich School of BioEngineering, Technische Universit\"at M\"unchen, 85748 Garching, Germany}

\author{F. Pfeiffer}
\affiliation{Lehrstuhl f\"ur Biomedizinische Physik, Physik-Department \& Munich School of BioEngineering, Technische Universit\"at M\"unchen, 85748 Garching, Germany}
\affiliation{Institut f\"ur Diagnostische und Interventionelle Radiographie, Klinikum rechts der Isar, Technische Universit\"at M\"unchen, 81675 M\"unchen, Germany}

\author{S. Karsch}
\email[E-mail: ]{stefan.karsch@mpq.mpg.de}
\affiliation{Ludwig-Maximilians-Universit\"at M\"unchen, Am Coulombwall 1, 85748 Garching, Germany}

\begin{abstract}
Laser wakefield acceleration of electrons represents a basis for several types of novel X-ray sources based on Thomson scattering or betatron radiation. The latter provides a high photon flux and a small source size, both being prerequisites for high-quality X-ray imaging. Furthermore, proof-of-principle experiments have demonstrated its application for tomographic imaging. So far this required several hours of acquisition time for a complete tomographic dataset. Based on improvements to the laser system, detectors and reconstruction algorithms, we were able to reduce this time for a full tomographic scan to 3 minutes. In this paper, we discuss these results and give a prospect to future imaging systems.
\end{abstract}


\maketitle

Laser-driven X-ray sources take the middle ground in brilliance and cost between low-cost microfocus X-ray tubes and large-scale synchrotron sources. This applies especially to sources based on laser-wakefield acceleration \cite{Esarey:2009ks}, which allow for the production of collimated, femtosecond X-ray beams \cite{Corde:2013bja}. In particular, Thomson backscattering and betatron radiation have proven to be the most relevant for applications in the hard X-ray regime \cite{Albert:2016do}. While the emission of radiation in both mechanisms is based on oscillatory motions of relativistic electrons, their oscillation frequencies differ significantly. Thomson backscattering sources rely on electrons oscillating in the electromagnetic field of an intense colliding laser pulse, whereas betatron radiation is generated by electrons wiggling transversely in the wake of a highly intense laser pulse traveling through a plasma while being accelerated \cite{Corde:2013bja}. Transverse electric fields of several GV/m to TV/m force the accelerated electrons with initial transverse momentum onto oscillating trajectories. This wiggler-like movement leads to a broadband X-ray emission, while the duration of the X-ray pulses is on the order of femtoseconds \cite{Phuoc:2007wm}.
Laser-driven sources are therefore particularly suitable to study ultrafast processes like transitions in the X-ray absorption near edge structures \cite{Saes:2003bn}. However, one of the most important medical applications of X-rays remains radiography. Single-shot X-ray imaging has been shown with both Thomson and betatron sources \cite{Fourmaux:2011cs,Dopp:2016ki} and tomographic imaging has been demonstrated using betatron sources \cite{Wenz:2015if,Cole:2016fh}. Moreover, the intrinsic small source size of a few microns allows for phase contrast imaging \cite{Kneip:2011cx,Najmudin:2014uu}.
While previous research focused on the potential of this imaging method, the acquisition time for a tomography in these studies has been on the order of several hours \cite{Wenz:2015if,Cole:2016fh}. Here we focus on the duration of such scans and report on a successful reduction of this time to a more application-relevant few minute scale. This was achieved by upgrading the experimental setup to support $1~\mathrm{Hz}$ repetition rates and making use of advanced reconstruction algorithms. The latter allows to acquire a single image per projection angle and perform a consistent reconstruction despite shot-to-shot X-ray flux fluctuations of the source. As a result, the data acquisition time for a centimeter-scale human bone sample was reduced from several hours to 180 seconds \cite{Dopp:2017ub}.

\begin{figure*}[ht]
	\centering
  \includegraphics[width=\textwidth]{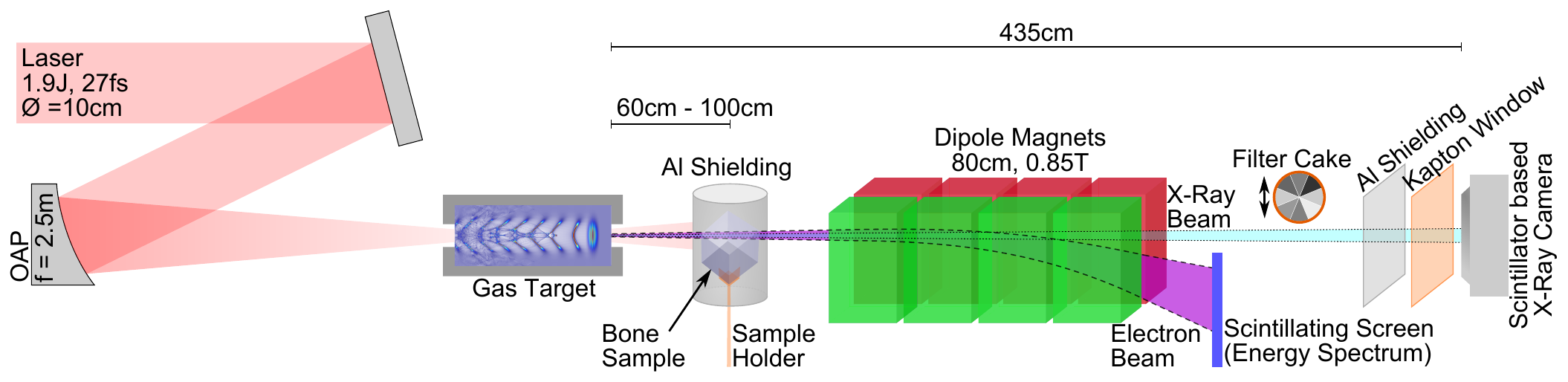}
	\caption{Experimental Setup for a tomography of a human bone sample at 1~Hz repetition rate. The laser pulse is focused into a gas cell where it ionizes hydrogen gas and drives a plasma wave. The resulting charge separation generates large electromagnetic fields. These accelerate electrons longitudinally while wiggling them transversely, which leads to betatron X-ray emission. The bone sample is protected from the laser light by a 15~$\upmu$m-thick aluminum foil which is passed by the electrons and X-rays. The electrons are deflected and analyzed in a magnetic spectrometer whereas the X-rays hit a scintillator which is imaged by a CCD. The camera itself is shielded by another 15~$\upmu$m thick aluminum foil. The geometric distances result in an image magnification of the bone sample of $\sim 4.4 - 7.3$ on the camera.}
\label{fig:fig1}
\end{figure*}

The measurements were performed at the Laboratory for Extreme Photonics in Garching, Germany. The $800~\mathrm{nm}$, $27~\mathrm{fs}$ laser pulse was delivered by the ATLAS (Ti:sapphire) laser system. An off-axis parabolic mirror (f/25) focuses the laser pulse containing an energy at the target position of $1.9~\pm0.1~\mathrm{J}$ to a spot size of $30~\upmu$m (FWHM intensity), which corresponds to a peak intensity of $5.5 \times 10^{18} {\mathrm{ W}}/{\mathrm{cm^2}}$ and a peak power of $70\pm4~\mathrm{TW}$ resulting in $a_0\approx1.6$.

As target a gas cell filled with hydrogen at a density of $\sim 5 \times 10^{18}~\mathrm{cm}^{-3}$ was used in which a movable piston allowed to adjust its length anywhere between 5 and 15 mm. The cell length was optimized with respect to the X-ray yield exploiting the fact that deep in the electron dephasing regime the electrons perform more betatron oscillations. The optimum X-ray yield was found at a cell length of 11~mm.

\begin{figure}%
	\centering
 \includegraphics[width=\linewidth]{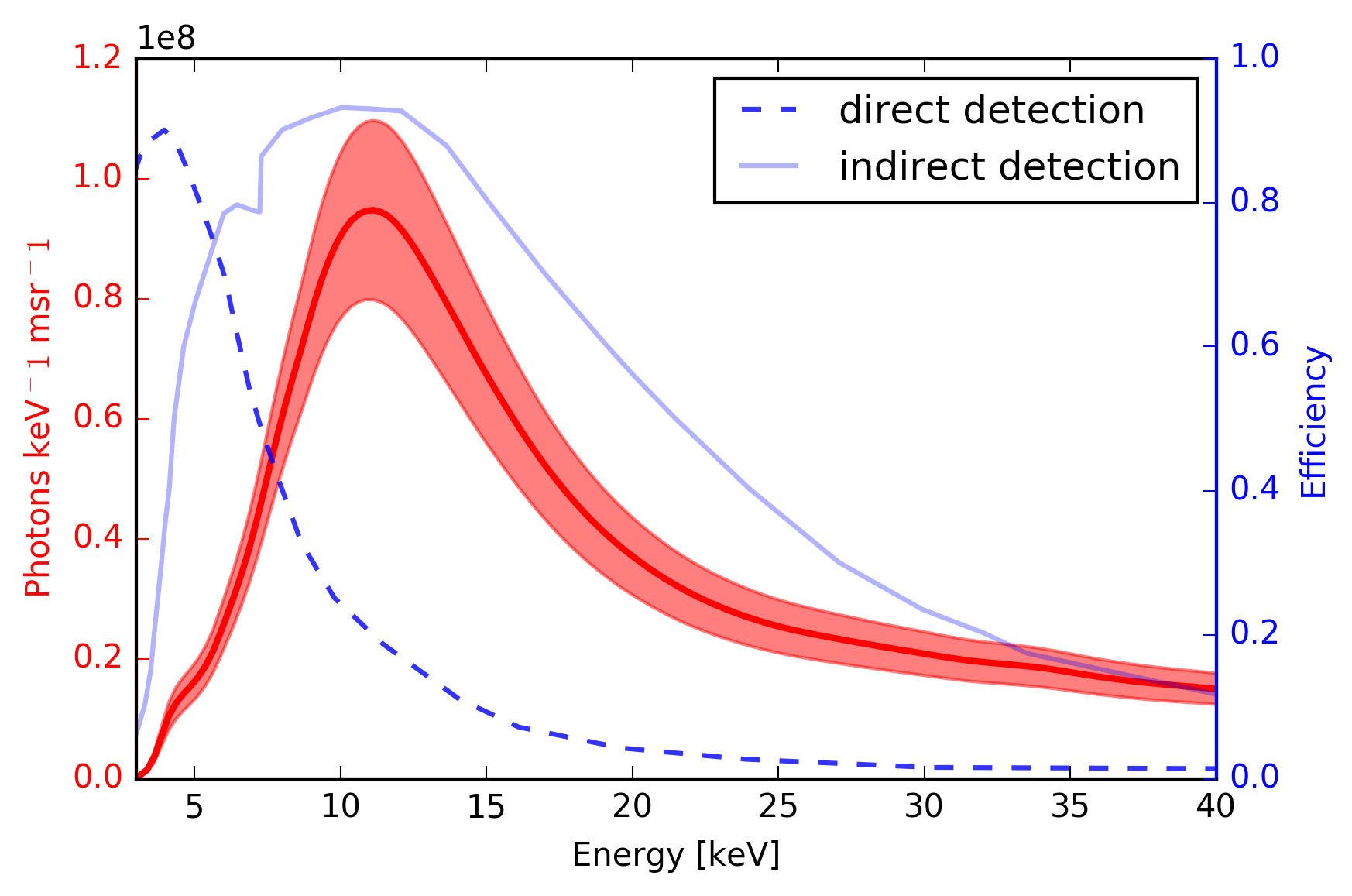}
	\caption{Comparison of the two different detection methods (blue) and reconstructed X-ray spectrum at the detector (red). The shaded red area indicates the corresponding rms error. At X-ray energies relevant for imaging bone samples, i.e. above 20~keV, indirect detection cameras have to be used due to their much higher sensitivity for high energetic X-rays.}
	\label{fig:qecams}
\end{figure}

The accelerated electrons are deflected onto a scintillating screen by a 0.8~m long 0.85~T dipole magnet of an electron spectrometer. We achieved an electron beam charge of $736\pm51$~pC with a pointing fluctuation of $1.1\pm0.1$~mrad. The X-rays are detected by a scintillator based camera located 4.35~m behind the gas cell exit. An array of aluminum filters of different thicknesses ranging from $5~\upmu$m to $610~\upmu$m can be inserted into the X-ray beam. The different transmission coefficients for each filter allow the determination of the X-ray spectrum by iteratively optimizing its calculated filter transmissions to the ones measured \cite{Sidky:2005eh}. The light shield and the Kapton vacuum window (cf. Fig.~\ref{fig:fig1}) are opaque for low energetic X-rays ($>50~\%$ transmission for energies above 7~keV) and therefore define a cut-off energy below which the spectrum cannot be reliably retrieved. Based on the spectral measurements, a time averaged flux up to $\left(1.6 \pm 0.35\right)\times 10^9$~photons/msr/s (behind the bone sample and light shield) at 1~Hz repetition rate is calculated. \label{sc} Comparing our retrieved spectrum to a synchrotron equivalent above the cut-off threshold of 7~keV gives a critical energy of $E_{\mathrm{crit}}=13.5 \pm 0.95~\mathrm{keV}$ (cf. Fig.~\ref{fig:qecams}). The size of the X-ray beam at the position of the CCD is larger than the sensor. Two dimensional Gaussian fits to the CCD images of the least divergent shots provide a lower estimate for the X-ray beam divergence of $12 \times 6~\mathrm{mrad}^2$ (s.d.).

To achieve a decent resolution, the sample was placed 60~cm to 100~cm behind the target which therefore was located in front of the electron beam spectrometer (in contrast to \cite{Cole:2016fh}). In order to minimize bremsstrahlung background from electrons colliding with the sample mount, the bone is supported by a thin 3D printed acrylic plastic holder \cite{Dopp:2016fa}. The signal of bremsstrahlung photons from the bone sample, holder and aluminum foil is negligible compared to the betatron signal. Furthermore, the sample is protected from laser radiation by an aluminum foil, which is wrapped around the rotation stage (and therefore rotates with the sample). 

In preceding experiments \cite{Wenz:2015if}, a direct detection X-ray CCD camera was employed for imaging purposes. This type of detector offers a high resolution (e.g. 13.5 $\upmu$m in the case of a Princeton Instruments \textit{PIXIS-XO:2048}), but its sensitivity drops rapidly for radiation beyond 10~keV. Furthermore, due to the high pixel number of the CCD chip and low-noise readout architecture, this type of cameras typically has a readout time of several seconds. This limits the shot frequency to around 0.1~Hz, which significantly increases the time needed for a tomography.

\begin{table}[htb]
\centering
  \begin{tabular}{ |l|cccccc| }
\hline
$E_\mathrm{X-ray} [\mathrm{keV}]$  & 5& 10 & 20 & 30   & 50& 80\\
\hline
$\chi_{\frac{1}{2}}$ [mm] & 0.02  & 0.13  & 1.8 & 2.7 & 8.5  &19.7\\
\hline

  \end{tabular}
  \caption{Half-value thickness $\chi_{\frac{1}{2}}$ for human bones at different energies \cite{akar2006measurement}.}
    \label{tab:hvt}
  \end{table}

In order to minimize the sample illumination - and therefore the exposure to ionizing radiation - the energy of the X-rays should be chosen such that the corresponding half-value thickness is on the order of the sample's physical thickness (cf. Table~\ref{tab:hvt}). This poses a major difficulty for medical imaging with direct-detection CCDs since they are insensitive to radiation transmitted by e.g. human bones. However, the detection efficiency for high energetic photons can be drastically increased by converting the incoming X-rays into photons of visible wavelengths via scintillators.

For the quick tomography experiments we therefore have used a scintillator-based CCD. It features an image intensifier with variable gain. This camera uses a \textit{Sony ICX285} Progressive Scan 2/3 rectangular CCD chip and supports frame rates up to 30 fps. In practice, the maximum shot frequency was limited to 1~Hz due to our laboratory data acquisition system. Due to the built in fiber-optics taper (50:11), the camera's effective pixel size is 29~$\upmu$m. The P43 phosphor scintillator offers sensitivity also for high X-ray energies (up to 100~keV) and therefore makes this indirect detection method suitable for medical imaging of thicker samples (cf. Fig.~\ref{fig:qecams}).

\begin{figure}
	\centering
  \includegraphics[width=\linewidth]{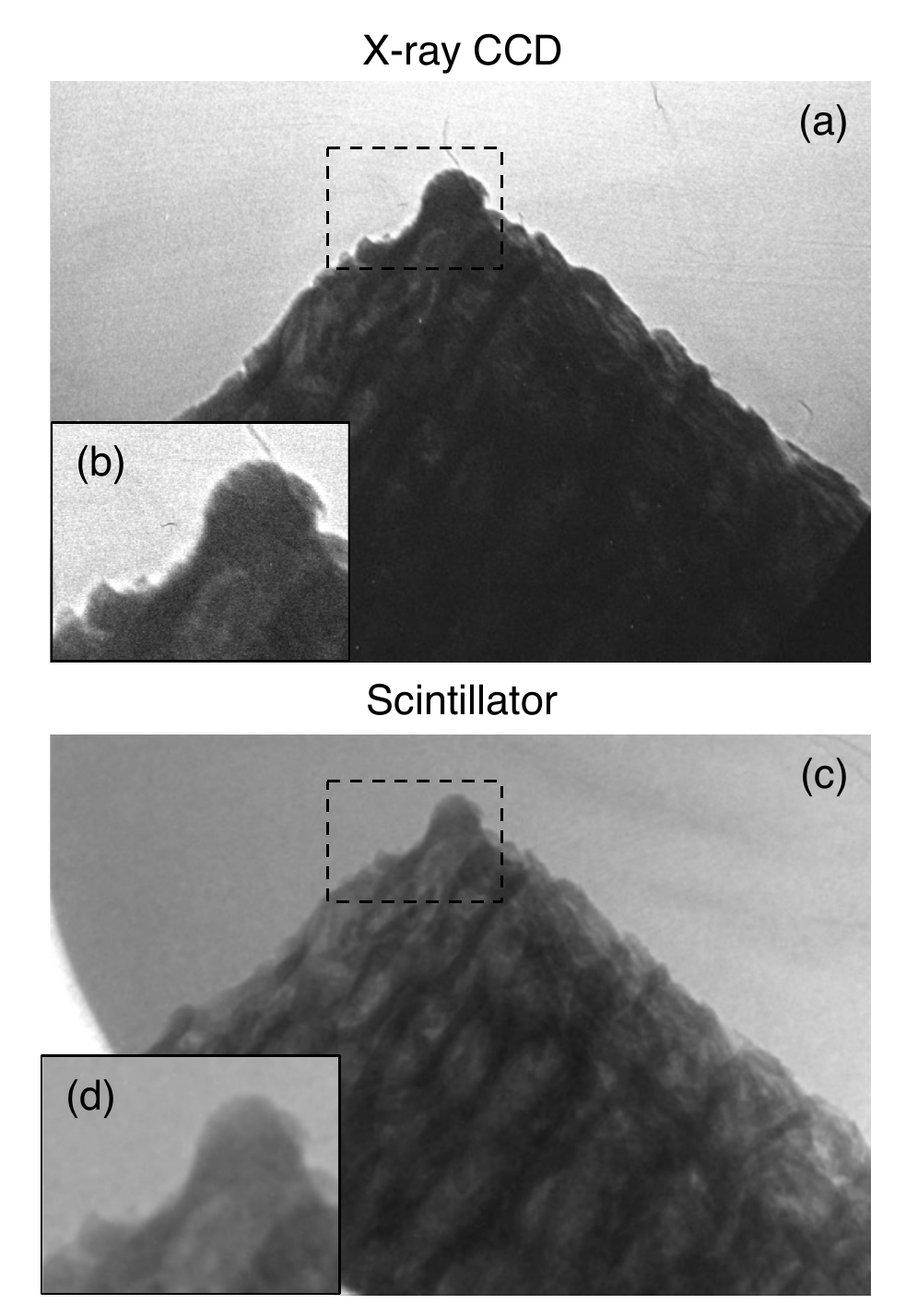}
	\caption{Comparison of imaging using an X-ray CCD camera and a scintillator-based detector. While the X-ray CCD produces sharper images, showing for instance signs of edge enhancement, the lower part of the bone sample remains almost opaque. In contrast, the scintillator camera is more sensitive in the $>10$~keV regime, clearly showing the trabecular structure.}
	\label{fig:fig2}
\end{figure}

Figure \ref{fig:fig2} shows a comparison for the two types of detectors at the same angle of the sample and similar X-ray spectrum. As shown in the insets, the direct-detection CCD has a better spatial resolution. In contrast, the photon energy range detected by the camera is not well transmitted by the bone and only limited information about its inner structure can be extracted. 

\begin{figure}[htb]
	\centering
  \includegraphics[width=\linewidth]{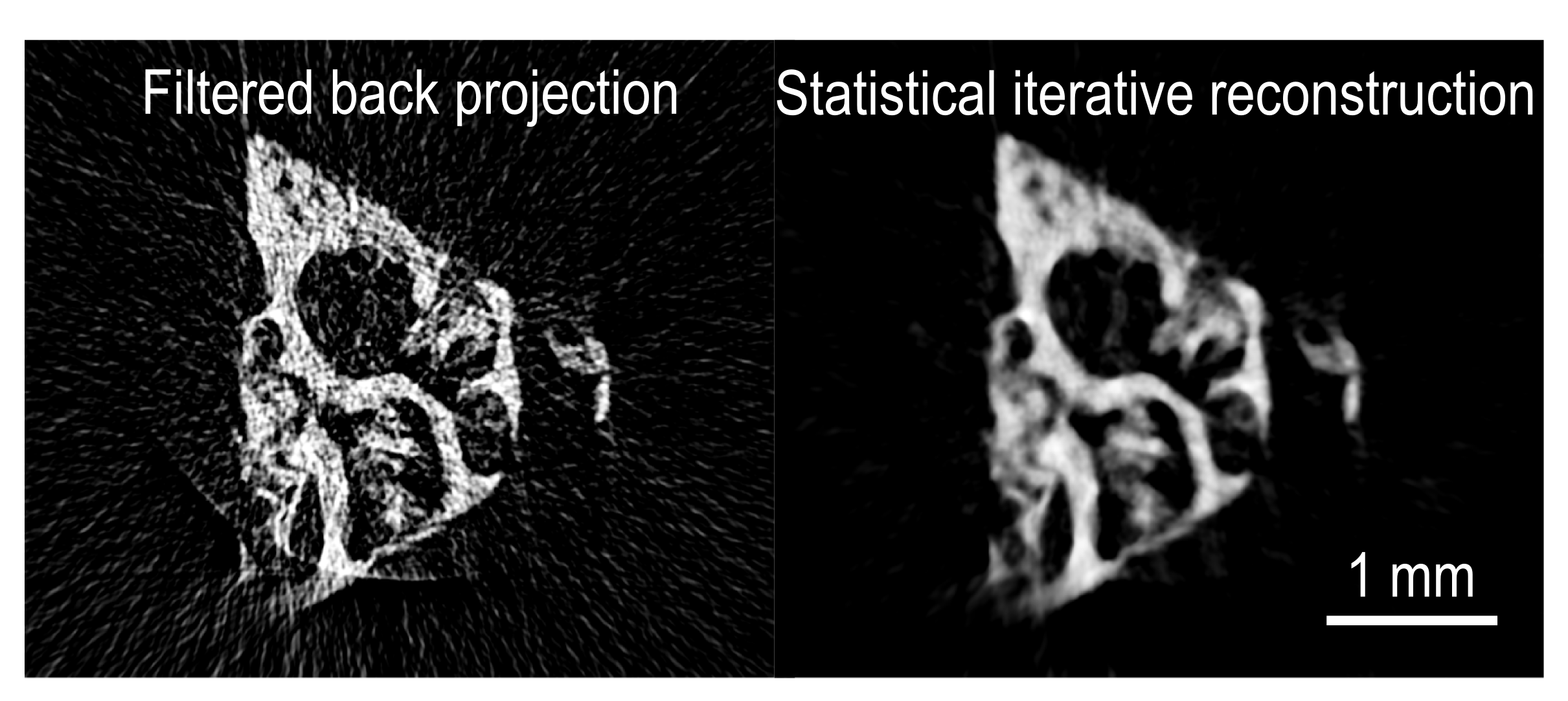}
	\caption{Quick tomography of a bone sample and comparison of two different reconstruction algorithms \cite{Dopp:2017ub}. In the left case filtered back projection (FBP) is used whereas the reconstruction in the right case was done via statistical iterative reconstruction (SIR). In contrast to FPB, SIR is capable of handling under-sampling artifacts which are due to the limited number of acquired tomograms.}
       \label{fig:cuts}
   \end{figure}

Another approach to reduce the acquisition time for medical applications is to minimize the necessary data underlying the tomographic reconstruction. This requires advanced reconstruction algorithms capable of handling reconstruction artifacts which are prone to appear for small data sets \cite{Dopp:2017ub}. The research of such algorithms has proven a prosperous field over the last two decades and lead to the development of statistical iterative algorithms (SIRs) which are now becoming state of the art for computed tomographic technology \cite{geyer2015state} (see Fig.~\ref{fig:cuts}). These algorithms iteratively improve the reconstructed sample and apply weighting factors to emphasize tomograms with better signal to noise ratio. As a proof of principle we performed a quick tomography encompassing 180 consecutive individual tomograms spanning $180^\circ$ at $1^\circ$ step size. The entire data set was taken within three minutes at 1~Hz repetition rate \cite{Dopp:2017ub}. 

The acquisition rate in these last experiments was limited by the data acquisition system, while both the vacuum and laser system would have supported shot frequencies up to 5~Hz. Beyond this, nowadays 100-TW-class laser systems with 10~Hz repetition rate are commercially available. If this could be fully exploited, a tomography as presented in this paper would take 18 seconds to acquire and a high-resolution tomography with 720 projection angles would only need slightly more than a minute - even with current laser technology. 

Nevertheless, demands on the photon energy continue to grow as full body CTs or non-destructive testing of thick samples require much higher photon energies, larger X-ray beam diameters and a higher time-averaged flux. Without any new mechanisms in the generation of betatron X-rays, the laser repetition rate must be increased in order to gain higher mean brilliances. However, this poses a challenging task which current commercially available laser systems cannot satisfy: conventional Ti:sapphire laser systems have typical average powers of $\sim50$ Watt which is limited by current crystal cooling concepts, such that an increased repetition rate usually comes at the cost of lower single-pulse energy. With new cooling concepts and/or laser architectures, this limit may be overcome in the future. Even though laser-wakefield acceleration at kHz repetition rates has recently been demonstrated using mJ-class laser systems \cite{Guenot:2017kv,Gustas:2018high,Salehi:2017bp}, the reduced pulse energy results in electron beams of lower energy (a few MeV) and negligible betatron emission. For the generation of high energetic X-rays with kHz repetition rates, sources based on Thomson-backscattering might be a viable alternative, since in this case the X-ray energy follows a favorable $E\propto 4\gamma^2$ scaling (e.g. 50~MeV electrons scattered with 800 nm light produce 60~keV radiation) \cite{Khrennikov:2015gxa}.

To conclude, we have demonstrated 1~Hz operation of a laser-driven betatron source for imaging applications. In the near-term, these betatron sources might be further improved by using controlled-injection schemes \cite{Dopp:2017dza} and operation at 10~Hz should be possible. At higher repetition rates it will be easier to reach the required X-ray energies for medical tomographies using Thomson backscattering sources. 

Based on the premise of laser-driven sources, the Munich universities LMU and TUM have established the new Centre for Advanced Laser Applications (CALA), which hosts several landmark laser installations aiming at both delivering higher laser pulse energies as well as providing a joule-scale kHz laser system. The former is met by the ATLAS-3000 laser, one of the few 1~Hz multi-petawatt laser systems in the world. The PFS-pro laser system in contrast will provide hundreds of mJ of laser pulse energy at kHz repetition rates. Both laser systems will drastically increase the average X-ray flux and therefore constitute another step towards real life imaging applications of laser-driven X-ray sources.

\section*{Acknowledgements}
The authors thank F. Schaff and T. Baum (TUM) for providing bone samples. This work was supported by DFG through the Cluster of Excellence Munich-Centre for Advanced Photonics (MAP EXC 158), the DFG Gottfried Wilhelm Leibniz program, TR-18 funding schemes, by EURATOM-IPP and the Max-Planck-Society.

\bibliographystyle{elsarticle-num}

\end{document}